\documentclass[a4paper,11pt]{article}

\usepackage[dvips]{graphicx}
\usepackage{amsmath,amssymb}
\usepackage{color}

\usepackage[normalem]{ulem}

\newcommand{\be} {\begin{equation}}
\newcommand{\ee}{\end{equation}}

\begin{document}

\author{Adam B.~Barrett*$^1$\\\\
\textit{$^{1}$ Sackler Centre for Consciousness Science} and \\
\textit{Department of Informatics}, University of Sussex, Brighton BN1 9QJ, UK\\
*adam.barrett@sussex.ac.uk (correspondence)\\}

\title{Stability of zero-growth economics analysed with a Minskyan model}

\date{}

\maketitle

\begin{abstract}
\noindent As humanity is becoming increasingly confronted by Earth's finite biophysical limits, there is increasing interest in questions about the stability and equitability of a zero-growth capitalist economy, most notably: if one maintains a positive interest rate for loans, can a zero-growth economy be stable? This question has been explored on a few different macroeconomic models, and both `yes' and `no' answers have been obtained. However, economies can become unstable whether or not there is ongoing underlying growth in productivity with which to sustain growth in output. Here we attempt, for the first time, to assess via a model the relative stability of growth versus no-growth scenarios. The model employed draws from Keen's model of the Minsky financial instability hypothesis. The analysis focuses on dynamics as opposed to equilibrium, and scenarios of growth and no-growth of output (GDP) are obtained by tweaking a productivity growth input parameter. We confirm that, with or without growth, there can be both stable and unstable scenarios. To maintain stability, firms must not change their debt levels or target debt levels too quickly. Further, according to the model, the wages share is higher for zero-growth scenarios, although there are more frequent substantial drops in employment.
\end{abstract}

\section{Introduction}
    As humanity is becoming increasingly confronted by Earth's finite biophysical limits, there is an increasing interest in questions about the stability and equitability of a zero-growth economy (Rezai and Stagl, 2016; Hardt and O'Neill, 2017; Richters and Siemoneit, 2017a). In particular, there has been a focus on the sustainability of a zero-growth economy that maintains a positive interest rate for loans. There are now a variety of models on which this question has been posed explicitly, and both `yes' (Berg et al., 2015; Jackson and Victor 2015; Rosenbaum, 2015; Cahen-Fourot and Lavoie, 2016) and `no' (Binswanger, 2009) answers have been obtained as to whether a stable zero-growth state is theoretically possible. Typically, the question is settled by the existence, or not, of a single attractive fixed point (i.e.~an equilibrium that is robust at least to small shocks) with economically desirable characteristics, namely positive profit and wage rates, and low unemployment (Richters and Siemoneit, 2017a). That is, the focus has been on demonstrating that there is some local stability within the system. However, real economies do not sit in equilibrium at a locally stable fixed point. They exhibit fluctuations, business cycles and, occasionally, severe crises, whether or not there is ongoing underlying growth in productivity with which to sustain growth in output (Minsky, 1986; Keen, 2011). This paper analyses zero-growth scenarios by focussing on global stability. Thus, a scenario is considered stable if its dynamics are characterised by fluctuations that do not grow in severity; unstable scenarios will be characterised by run-away explosive behaviour (which would correspond to a crisis). The model employed is a non-linear dynamical system that incorporates elements of Minsky's financial instability hypothesis (FIH) (Minsky, 1986; Minsky, 1992). The analysis involves the tweaking of a productivity growth parameter, set to either two percent or zero to respectively produce growth and no-growth scenarios. In so doing, this paper is the first to attempt to compare the relative stability of a zero-growth economy with that of a growing economy.

Key to the FIH is that serious macroeconomic instability arises as a result of firms desiring to vary their debt burden in response to changes in the profit share, and expectations about the future profit share. This idea was first put into a mathematical model by Keen (1995), and there is now a substantial literature on Minskyan models that capture various dynamics related to the FIH; see Nikolaidi and Stockhammer (2017) for a recent survey. The original Keen (1995) model consisted of three coupled differential equations for the key variables: wage rate, employment rate and firm debt. It is derived from a few simple intuitive assumptions, and is capable of producing both stable and unstable scenarios, depending on firms' behaviour in relation to debt. It thus provides a useful starting point from which to build a simple model to compare the stability of growth and no-growth scenarios. Further, the presence of labour dynamics (\textit{a la} Goodwin) enables comparison of employment and wage rates between growth and no-growth scenarios. However, in the original model, investment is a direct function only of the profit share of output, i.e.~investment decisions are based purely on recent profit. Since investment must depend on growth, it is necessary for the present study to extend the model. Further, it is realistic for investment decisions to have an additional explicit direct dependence on debt, (i.e.~beyond the indirect dependence due merely to profit itself depending on debt). Thus, rather than employing the Keen (1995) model in its original form, the investment dynamics here have terms added from a recent model of Dafermos (2017) to include an explicit direct dependence on growth and debt.\footnote{Running the original Keen (1995) model with parameters that produce a stable scenario with two percent productivity growth led to run-away behaviour when productivity growth was switched to zero (simulation not shown). See Appendix \ref{sec:generality} for an explanation of this, and further discussion of the original Keen model.}  With output determined by the investment dynamics, consumption will be the accommodating, or residual, variable in the model.

The outline of the paper is as follows. Section \ref{sec:model} presents the details of the model. The dynamical variables in the model are the wage rate, the employment rate, firm debt and target firm debt. Further equations express GDP, growth rate and profit share in terms of these variables. In Section \ref{sec:equilibrium} the analysis pipeline is presented. Section \ref{sec:SFC} demonstrates that the modelled dynamics can form part of a stock-flow consistent framework. In Section \ref{sec:parameters} the parameters used in the simulations are written down and explained. Section \ref{sec:results} presents the simulation results. Scenarios of constant positive productivity growth and constant zero productivity growth are shown, demonstrating stable and unstable runs for both cases. Then, more realistic scenarios of fluctuating productivity growth are explored, with comparisons between scenarios in which mean growth is positive and in which mean growth is zero. Further, transitions from a positive to zero productivity growth era are considered. The paper concludes with Discussion and Concluding Remarks sections.

 \section{The Model} \label{sec:model}

This section describes the model and its assumptions in detail. As mentioned in the Introduction, most of the pieces of the model are taken from that of Keen (1995), but the debt dynamics are inspired by the recent model of Dafermos (2017). The notation and presentation are drawn from Grasselli and Costa Lima (2012). Further, the model is an extension of the Goodwin (1967) growth cycle model, which consisted of just two equations for the wage and employment rates, and contained no debt, only reinvestment of profit. 

It is assumed that there is full capital utilisation and a constant rate of return $\nu^{-1}$ on capital $K$:
\be
Y=K/\nu\,=aL, \label{eq:output}
\ee
where $Y$ is the yearly output, $a$ is productivity and $L$ is labour employed. The yearly wage bill is denoted $W$, firm debt is denoted $D$, and the interest rate by $r$. The yearly profit $\Pi$ is defined as output minus the yearly wage bill minus the yearly interest payments, that is $\Pi=:Y-W-rD$. Concerning investment, it is assumed that all profits are either reinvested or used to pay down debts. Thus the rate of investment $I$ is given by\footnote{The dot here denotes derivative with respect to time. Note the continuous time formulation implies that profit and investment here are both rates. The term yearly profit is used in place of profit rate to avoid confusion, as profit rate commonly refers to a rate of return on capital.}
\be
I=\dot{D}+\Pi\,.  \label{eq:investment}
\ee
This is admittedly a simple model of finance, however the concern in this paper is to construct just one possible economic model with interest-bearing debt and no growth imperative; for further discussion of finance see Section \ref{sec:SFC} and the Discussion. Given the rate of depreciation of capital $\delta$ we have
\be
\dot{K}=I-\delta K\,. \label{eq:capitaldot}
\ee
From \eqref{eq:output}, \eqref{eq:investment} and \eqref{eq:capitaldot} we have
\be
\dot{Y}=\frac{1}{\nu}(\dot{D}+\Pi - \delta K )\,, \label{eq:YDK}
\ee
an expression we will use further down to derive the growth rate in terms of profit and debt. Productivity growth is denoted by $\alpha$, and a constant population size $N$ is assumed. Thus
\be
\dot{a}=\alpha a\,. \label{eq:prodgrowth}
\ee
Using \eqref{eq:output} and \eqref{eq:prodgrowth} it can be derived that the employment rate $\lambda=:L/N$ satisfies
\be
\dot{\lambda}=\lambda(g-\alpha)\,, \label{eq:lambda}
\ee
where $g=:\dot{Y}/Y$ is growth (of output). The rate of change of wages $w$ per unit of labour is an increasing function of the employment rate $\lambda$,
\be
\dot{w}=\Phi(\lambda)w\,, \label{eq:wdot}
\ee
reflecting the assumption that the higher the rate of employment, the greater the bargaining power of workers. We specify the Phillips curve\footnote{Throughout the paper, the term `Phillips curve' refers to that linking the rate of employment with wage growth, rather than that linking wage growth and inflation.} $\Phi$ explicitly in Section \ref{sec:parameters} below. Note that in addition to being an increasing function, the Phillips curve should satisfy $\Phi(0)<0$ to ensure there is an employment rate below which there is downward pressure on wages. Further, the curve should rise steeply as $\lambda$ approaches 1 from below, as the employment rate cannot rise higher than 1 (given that it starts positive, Eq.~\eqref{eq:lambda} ensures that it can't drop below zero). In practice, in the simulations, an exceptional line was included in the code to implement that if $\lambda$ exceeds 0.99, and Eq.~\eqref{eq:lambda} indicates that $\lambda$ should rise further, then that equation is overridden, and $\dot{\lambda}$ is set to zero for the given integration step. This is just a simple way of imposing that there is a limited labour pool. 

The equation for the wages share of output $\omega=:wL/Y$ is derived from \eqref{eq:output}, \eqref{eq:prodgrowth} and \eqref{eq:wdot} as
\be
\dot{\omega}=\omega[\Phi(\lambda)-\alpha]\,. \label{eq:omega}
\ee
The equations \eqref{eq:lambda} and \eqref{eq:omega} for the employment rate and wages share are the same as those of the Goodwin (1967) model, except growth $g$ itself satisfies different dynamics in the present model, as will be described below.

Considering now the debt dynamics, following Dafermos (2017), the rate of change of debt is taken to be proportional to the difference between the target debt and the current debt. The equation for this, expressed in terms of normalised debt $d=:D/Y$ is
\be
\dot{d}=\theta_1(d_T-d)\, \label{eq:ddot}
\ee
(henceforth, when the term debt is used, normalised debt is implied).\footnote{This equation implies that non-normalised debt $D$ satisfies $\dot{D}=\theta_1(d_T-d)Y+dg$.  Thus, it is assumed that the rate of increase of debt depends not just on how far away the current stock of debt is from the current target, but also on the current growth rate of the economy, so as to achieve the desired move of the debt-to-output ratio toward the target.} The parameter $\theta_1$ here determines the rate at which debt moves towards the target level; $\theta_1^{-1}$ is the length of time it takes for the difference between debt and target debt to drop by a factor of $e$, all other variables remaining constant. Note that, in practice, target debt may never become close to being realised, as all the variables of the system remain in continuous flux. The target debt has a tendency to move towards a benchmark that depends on the current growth rate and profit share $\pi=:\Pi/Y$:
\be
\dot{d}_T=\theta_2(d_0+\eta_1 g + \eta_2 \pi -d_T)\,. \label{eq:dT}
\ee
The parameter $\theta_2$ determines the timescale on which target debt moves towards the benchmark $d_0+\eta_1 g + \eta_2 \pi$. The parameter $d_0$ is a constant, and $\eta_1$ and $\eta_2$ respectively determine how strongly the benchmark debt is affected by changes in growth rate and profit share. As mentioned above, in the original Keen (1995) model, the investment rate was taken as a function only of profit, with the simplifying assumption that firms pay attention only to profits and not to debt at all. The target debt equation \eqref{eq:dT} differs from that in Dafermos (2017) by depending additionally on the profit share as well as the growth rate. Note that these dynamics are designed to model `normal times', and the onset of a crisis, but not the behaviour of the economy after crisis onset. A crisis is assumed to have occurred if at any point in a simulation, investment becomes less than zero as a result of the change of debt becoming sufficiently negative. 

Given the above, the final set of equations for the model can be written down. The non-redundant dynamical variables are wages share $\omega$, employment rate $\lambda$, debt $d$ and target debt $d_T$. The profit share $\pi$, growth rate $g$ and output $Y$ (GDP) can be written in terms of these variables. The profit share is given by
\be
\pi=1-\omega-r d\,, \label{eq:profitshare}
\ee
where $r$ is the interest rate on loans. Using \eqref{eq:output}, \eqref{eq:YDK}, \eqref{eq:ddot}, and some basic calculus,\footnote{Specifically the product rule $\dot{D}=\dot{d}Y+\dot{Y}d$.} the growth rate can be expressed as
\be
g=\frac{\pi+\theta_1(d_T-d)-\delta\nu}{\nu-d}\,. \label{eq:growtheqn}
\ee
The output derives, by definition and basic calculus, from the integral of the growth rate
\be
Y(t)=Y_0\,\mathrm{exp} \left( \int_{t_0}^{t} g\, \mathrm{d} t \right)\,,
\ee
where $Y_0$ is output at some initial time $t_0$. Finally, the four coupled differential equations that specify the dynamics of the system are equations \eqref{eq:dT}, \eqref{eq:ddot}, \eqref{eq:lambda} and \eqref{eq:omega}:
\begin{eqnarray}
\dot{d}_T&=&\theta_2(d_0+\eta_1 g+ \eta_2 \pi -d_T) \label{eq:system1}\\
\dot{d}&=&\theta_1(d_T-d)\\
\dot{\lambda}&=&\lambda(g-\alpha)\\
\dot{\omega}&=&\omega[\Phi(\lambda)-\alpha]\,. \label{eq:system4}
\end{eqnarray}

\subsection{Analysis pipeline, fixed point and instability} \label{sec:equilibrium}
As mentioned in the Introduction, recent analyses of zero-growth economics have focused on the fixed points (equilibria) of the model systems (Richters and Siemoneit, 2017a). Here however, the focus is on non-equilibrium dynamics. The system will be classed as stable if it is not prone to a crisis (characterised by explosive run-away behaviours), even if it does not converge to a fixed point. The justification for this is that real economies are constantly fluctuating and exhibit oscillations; it is unrealistic to expect convergence to a fixed point.

In practice the system tends to oscillate around the theoretical fixed point, and thus it remains informative to write down the equations for it. There exists one economically desirable fixed point, i.e.~one with a positive employment rate $\lambda>0$ and positive wages share $\omega>0$. Setting the left-hand side of each of the equations \eqref{eq:system1}--\eqref{eq:system4} to zero, assuming $\lambda>0$ and $\omega>0$ and using also \eqref{eq:growtheqn} and \eqref{eq:profitshare}, the fixed point can be derived as being given by
\begin{eqnarray}
\bar{\lambda}&=&\Phi^{-1}(\alpha)\,, \label{eq:equil1}\\
\bar{d}_T=\bar{d}&=&\frac{1}{1+\eta_2\alpha}[ d_0+\eta_1\alpha+\eta_2\nu(\delta+\alpha)]\,,\\
\bar{\omega}&=&1-(\alpha+\delta)\nu-(r-\alpha)\bar{d}\,,\\
\bar{\pi}&=&\delta\nu+\alpha(\nu-\bar{d})\,, \\
\bar{g}&=&\alpha\,.  \label{eq:equil5}
\end{eqnarray}
The nature of the fixed point, i.e., whether it is attractive or repulsive, can be formally assessed by the signs of the eigenvalues of the Jacobian matrix, see Appendix \ref{sec:fixedpoint} for details. There is no simple set of conditions on the parameters for the fixed point to be attractive. In the scenarios carried out below (Section \ref{sec:results}), there are cases for which the fixed point is attractive and cases for which it is repulsive.

In the simulations, the system is started close to, but not at, the fixed point. When all parameters are held constant, three possible behaviours are exhibited (i) convergence to the fixed point; (ii) oscillations around the fixed point, with an amplitude that eventually stabilises; (iii) oscillations that grow in amplitude, until a crisis is reached and the simulation is stopped (investment becomes negative and the model breaks down). Scenarios exhibiting either of the first two behaviours are considered stable scenarios, whilst only the third scenario is considered unstable. Further scenarios are considered in which productivity growth fluctuates randomly around a fixed mean. In these scenarios, the nature of the fixed point becomes irrelevant, as the fluctuations in productivity growth trigger ongoing oscillations in all the variables whether the fixed point is attractive or repulsive. This analysis pipeline differs from that employed by Richters and Siemoneit (2017a), which classified a model scenario as stable if and only if there exists an attractive fixed point with desirable economic characteristics.

\subsection{Stock-flow consistency} \label{sec:SFC}

\begin{table}
\begin{center}
\caption{A financial balance sheet for the model}
\footnotesize{
\begin{tabular}{c|c|c|c|c|c}
\hline
 &Households & Firms  & Banks  & Foreign & $\Sigma$  \\
 \hline
 Net financial assets & $S$ & $-D$ & $D-S-F$ & $F$ & 0 \\
 Financial assets & $S$ & - & $D$ & $F$ & $S+D+F$ \\
 Deposits & $S$ & - & - & $F$ & $S+F$\\
 Loans & - & - & $D$ & - & $D$\\
 Financial liabilities & - & $D$ & $S+F$ & - & $D+S+F$ \\
 Deposits &- &- & $S+F$ & - & $S+F$\\
 Loans & - & $D$ & - & - & $D$ \\
 \hline
 \end{tabular} \label{tab:flows}
}
\end{center}
\end{table}

\begin{table}
\begin{center}
\caption{A transaction flow matrix for the model}
\footnotesize{
\begin{tabular}{c|c|cc|cc|c|c}
\hline
 &Households & Firms  & & Banks  & & Foreign & $\Sigma$  \\
 &                              & Current & Capital &           Current & Capital     &                 & \\
\hline
Wages & $\omega Y$ & $-\omega Y $&&& & & 0 \\
Consumption & $-C$ & $C-M_C$ && && $M_C$ &  0\\
Investment & & $I-M_I$&-I & && $M_I$&  0 \\
Exports & & $X$ && & &$-X$ & 0\\
Interest on loans &  & $-rD$& &$rD$& & & 0 \\
Interest on deposits & $i_S$ && & $-i_S$ && & 0\\
Interest to RoW & & & & $-i_F$ && $i_F$ & 0\\
Bank profits & $rD-i_S-i_F$ && & $-(rD-i_S-i_F)$ && & 0\\
Firm profits & & $-\Pi$ & $\Pi$ & & & & 0\\
Net new loans & & & $\dot{D}$ & & -$\dot{D}$ & & 0 \\
New savings & $-\dot{S}$ &&&& $\dot{S}$ & & 0\\ 
New debt to RoW & & & & & $\dot{F}$ & $-\dot{F}$ & 0\\
\hline
$\Sigma$ & 0 & 0 & 0 & 0 & 0 & 0 & 0\\
\hline
\end{tabular} \label{tab:flows}
}
\end{center}
\end{table}

In this section it is demonstrated that the model can fit into a stock-flow consistent framework. Table 1 provides a financial balance sheet for the model, and Table 2 provides a transaction flow matrix consistent with the model. Note that not all the flows in the transaction flow matrix are specified explicitly in the equations of the model, and that Tables 1 and 2 do not provide the unique stock-flow consistent framework that is compatible with the model. It rather provides a useful simple example framework for conceptualising the model, and demonstrating its consistency. It is assumed that the flows that are not specified explicitly do not affect the long run stability of the system. The financial assets are household savings $S$, firm debt $D$, and net debt $F$ to the rest of the world (RoW). The flows that have not been defined in the previous section are domestic household consumption $C$, interest on savings $i_S$, interest to the RoW $i_F$, imports $M$, sub-divided into those for consumption $M_C$ and those for investment $M_I$, and exports $X$. Note that domestic consumption, imports and exports must satisfy the accounting identity
\be
Y=C+I+X-M\,. \label{eq:GDP}
\ee

The lack of a role for consumption in the stability of the model constitutes a departure from several recent analyses of zero-growth economics (Richters and Siemoneit, 2017a). However, those analyses assumed constant rates of consumption out of wealth and income. This is reasonable for their fixed point analyses, which explore the system only in the immediate neighbourhood of the fixed point. However, for the dynamical analyses in the present study, including an explicit role for consumption would involve adding further parameters to the model to specify how consumption rates depend on all of the dynamical variables, particularly on the current growth rate and wages share. Further assumptions would have to be made about the availability of credit to households for consumption beyond income. It is beyond the scope of this study to consider diverse debt behaviour for households; it is rather assumed that if households are not overly indebted, then consumption does not play a role in stability of zero growth macroeconomics. Thus, for the present study the role of consumption is ignored. The addition of a foreign sector, capable in theory of smoothly consuming output that is not consumed domestically makes this assumption more reasonable than if the foreign sector were excluded. It is left for future work to introduce roles for domestic consumption, imports and exports in this modelling framework.

Note that the equations of the model impose that all investment is financed by firm profit and (domestic) bank lending, rather than through households or banks taking up firm equity, or households lending to firms. It is further assumed here that banks distribute all their profits to households. The model also neglects to include financial speculation. Incorporating explicit details of realistic modern-day finance into the model is left for future work, see Discussion. 

\subsection{Parameter values} \label{sec:parameters}
For the constants in the model, typical values are chosen, taken from Jackson and Victor (2015). The interest rate on loans is $r=0.05$, i.e.~5\%. The depreciation rate is 
$\delta=0.07$, since typical values in advanced economies are around 6-8\%. The capital to income ratio is $\nu=3$; the current value for this in Canada is a little under 3, while in the UK the value for this is around 5. The Phillips curve $\Phi$ is drawn from Keen (2013) and is given by
\be
\Phi(\lambda)=0.01\mathrm{exp}[50(\lambda-0.95)]-0.01\,,
\ee
so that $\Phi(0.95)=0$ and $\Phi(0)\approx -0.01$. Note that at the fixed point \eqref{eq:equil1}--\eqref{eq:equil5} only the employment rate depends on the Phillips curve. In particular, neither the profit or wages share at the fixed point depend on the Phillips curve.

\section{Simulation scenarios} \label{sec:results}

This section presents the results of the simulations. Scenarios with constant positive and zero productivity growth are explored, as well as scenarios in which productivity growth fluctuates, and in which there is a transition from positive to zero productivity growth. Further, the dependence of stability on the debt behaviour parameters $\theta_1$, $\theta_2$, $\eta_1$, $\eta_2$ and $d_0$ is investigated. Stability is assessed based on whether or not a crisis occurs, where a crisis is defined as occurring if a moment is reached at which investment turns negative as a result of rapid debt pay-off. When a crisis occurs, the model is assumed to have broken down, and the simulation is halted. 

 \begin{figure*}
 \vspace{-2cm}
\begin{center}
\includegraphics[width=0.99\textwidth]{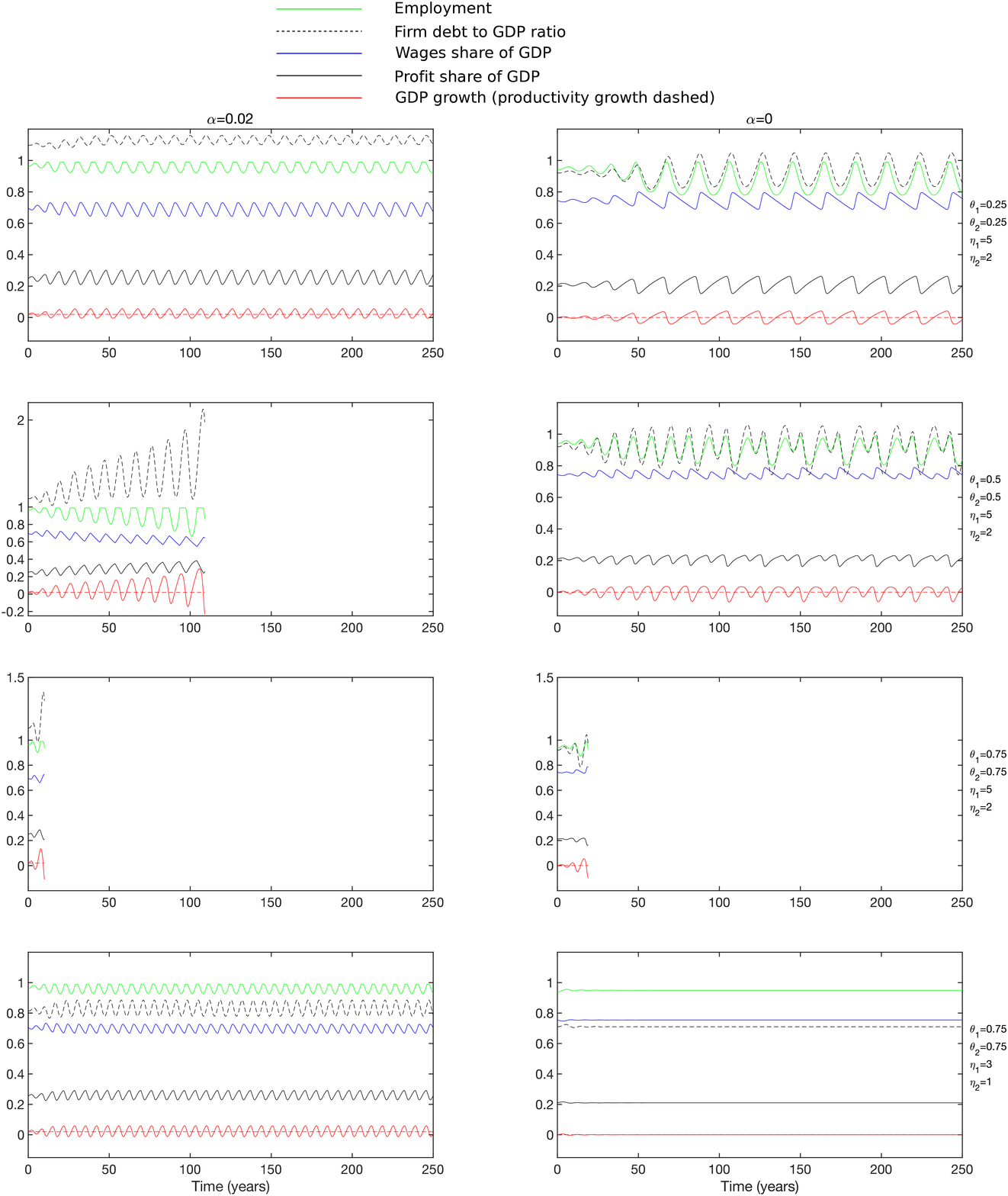}
\end{center}
\caption{Example two percent (left) and zero (right) constant productivity growth runs for different debt behaviour parameters. In the top row $\theta_1=\theta_2=0.25$, $\eta_1=5$, $\eta_2=2$. In the second row, $\theta_1$ and $\theta_2$ are increased to 0.5, leading to instability for the $\alpha=0.02$ case. In the third row, $\theta_1$ and $\theta_2$ are increased to 0.75, leading to instability for both the positive and zero growth cases. The fourth row shows stability of positive and zero growth cases for  $\theta_1=\theta_2=0.75$, $\eta_1=3$, $\eta_2=1$. In each panel $d_0=0.5$. All variables are started at the value they take at the fixed point, except for $\lambda$ which is initialised at $\bar{\lambda}-0.01$.} \label{fig:profitvolcases}
\end{figure*}

Fig.~\ref{fig:profitvolcases} shows two percent\footnote{This was a typical value for advanced countries during the economically stable period 1981-2006; see OECD data at https://data.oecd.org.} and zero constant productivity growth scenarios for several choices of the debt behaviour parameters. In the top row, the strength of dependence of benchmark debt on current growth and profit share are respectively $\eta_1=5$, $\eta_2=2$, while the rates of convergence of debt to target debt and target debt to benchmark debt are given by $\theta_1=\theta_2=0.25$, corresponding to a timescale of 4 years (for an $e$-fold convergence). The constant $d_0=0.5$. This leads to actual debt ratios in the various scenarios presented in Fig.~\ref{fig:profitvolcases}  lying in the same range as those currently typical in advanced economies.\footnote{As obtained from the OECD's table entitled `Debt of non-financial corporations, as a percentage of GDP'. Available from http://stats.oecd.org/index.aspx?queryid=34814\#.} Simulations are initialised with all variables assigned the values they take at the fixed point except for the employment rate $\lambda$, which is initialised at its fixed point value minus 0.01, so as to avoid a constant equilibrium scenario. It can be seen that for these parameter choices the system is stable for both positive and zero productivity growth, although the zero growth case exhibits higher fluctuations in employment. GDP growth fluctuates close to productivity growth, as one would expect, given that a constant population size is assumed. Note that 250 years was sufficient to display the behaviour of these and all other subsequent parameter choices. Continuing the simulation for longer merely resulted in repetitive oscillatory behaviour. In the second row of Fig.~\ref{fig:profitvolcases}, the debt change parameters $\theta_1$ and $\theta_2$ are both increased to 0.5, corresponding to a timescale of 2 years for ($e$-fold) convergence of debt to target debt and target debt to benchmark debt. This led to a crisis occurring during the two percent productivity growth run, while the zero growth run remained stable, albeit with oscillations. In the third row, $\theta_1$ and $\theta_2$ are increased further to 0.75, and a crisis occurs for both positive and zero growth cases. Finally, in the bottom row of Fig.~\ref{fig:profitvolcases}, $\theta_1$ and $\theta_2$ are maintained at 0.75, while $\eta_1$ and $\eta_2$ are reduced, respectively to 3 and 1. This leads again to a stable outcome for both two percent and zero productivity growth. The unstable scenario in the left panel of the second row, i.e.~two percent productivity growth, $\theta_1=\theta_2=0.5$, $\eta_1=5$, $\eta_2=2$, could be rendered stable by decreasing any one of the debt behaviour parameters, e.g.~by changing either $\theta_1$ to 0.25, $\theta_2$ to 0.25, $\eta_1$ to 3 or $\eta_2$ to 1, or from reducing the constant $d_0$ to 0.3. In general, the system has potential to move from being stable to unstable if any of the debt behaviour parameters $\theta_1$, $\theta_2$, $\eta_1$, $\eta_2$ and $d_0$ are increased from a given stable scenario. In summary, Fig.~\ref{fig:profitvolcases} demonstrates that the model allows for both stable and unstable economic scenarios, and, in concordance with Minsky (1986, 1992), the greater the variability in debt, the more likely the scenario ends in crisis. The model can be stable for zero productivity growth as well as for positive productivity growth, and we have even found a scenario in which the model is stable for zero but not two percent productivity growth.

The model assumes a constant interest rate. Incorporating a dynamical interest rate, as well as interactions between the interest rate and the other parameters and variables of the model is beyond the scope of this paper. However, if everything else is held constant, and the interest rate is increased, the dynamics generally become more stable. Fig.~\ref{fig:figappinterest} in Appendix \ref{app:figs} shows an example of this, namely by reproducing the scenarios in the second row of Fig.~\ref{fig:profitvolcases} but with a higher interest rate of $r=0.1$. With this higher interest rate, the $\alpha=0.02$ case no longer exhibits a crisis, and the size of the oscillations for the $\alpha=0$ case are smaller than for the lower interest rate of $r=0.05$. Note though that, in the model, debt decisions have no direct dependence on the interest rate (\ref{eq:ddot}, \ref{eq:dT}). If, more realistically, there were a term in the equation \eqref{eq:ddot} for the rate of change of debt that was proportional to the interest rate, then the contribution of this term to the overall debt volatility would increase for a higher interest rate. This would have a tendency to make the model less stable for a higher interest rate.

\begin{figure*}
\begin{center}
\includegraphics[width=0.96\textwidth]{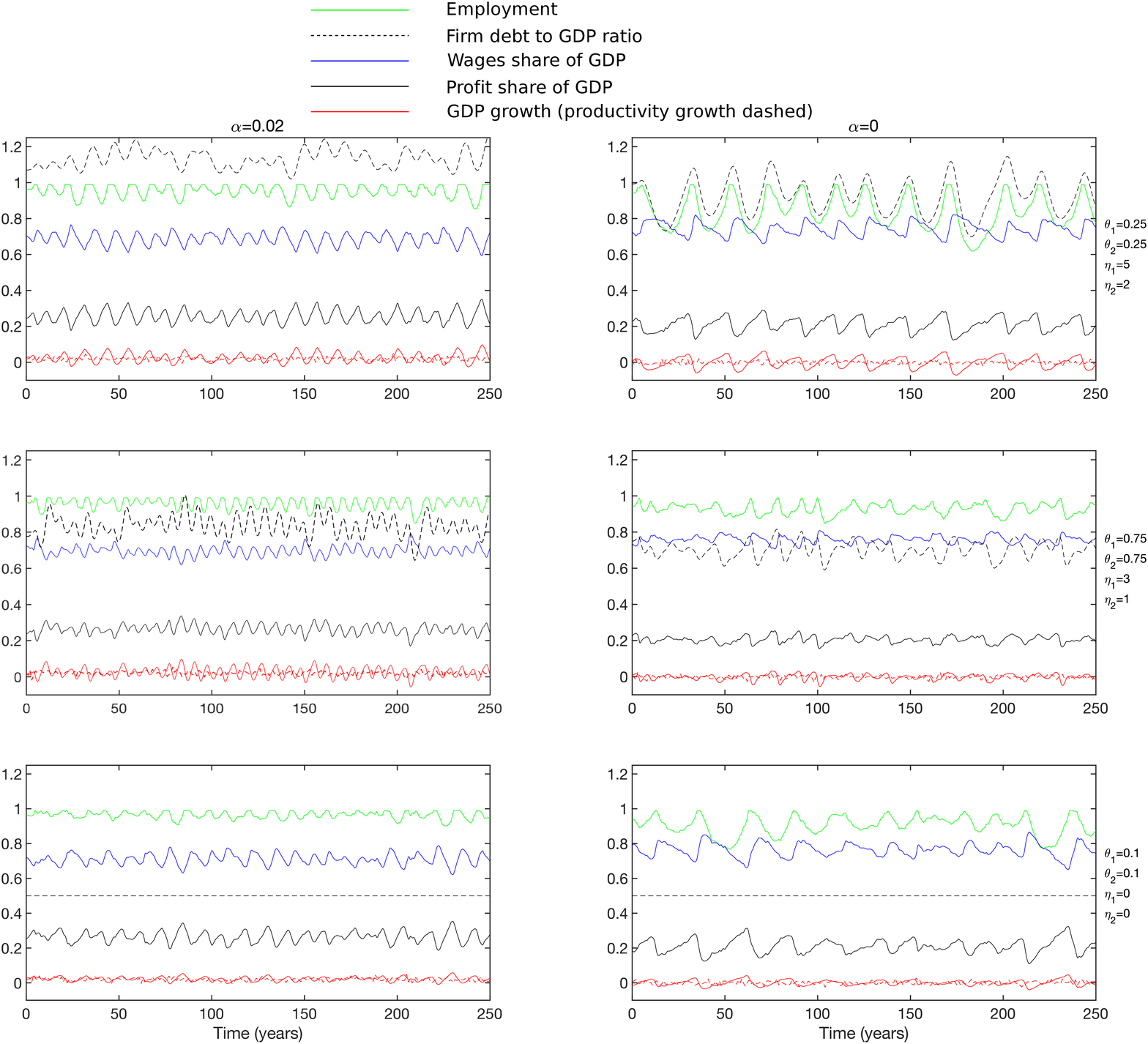}
\end{center}
\caption{Stochastic productivity growth runs. (Left) Two percent mean productivity growth. (Right) Zero mean productivity growth. In all panels $d_0=0.5$. All variables are started at the value they take at the fixed point, except for $\lambda$ which is initialised at $\bar{\lambda}-0.01$. See main text for further details.} \label{fig:stochcases}
\end{figure*}

\begin{figure*}
\begin{center}
\includegraphics[width=0.96\textwidth]{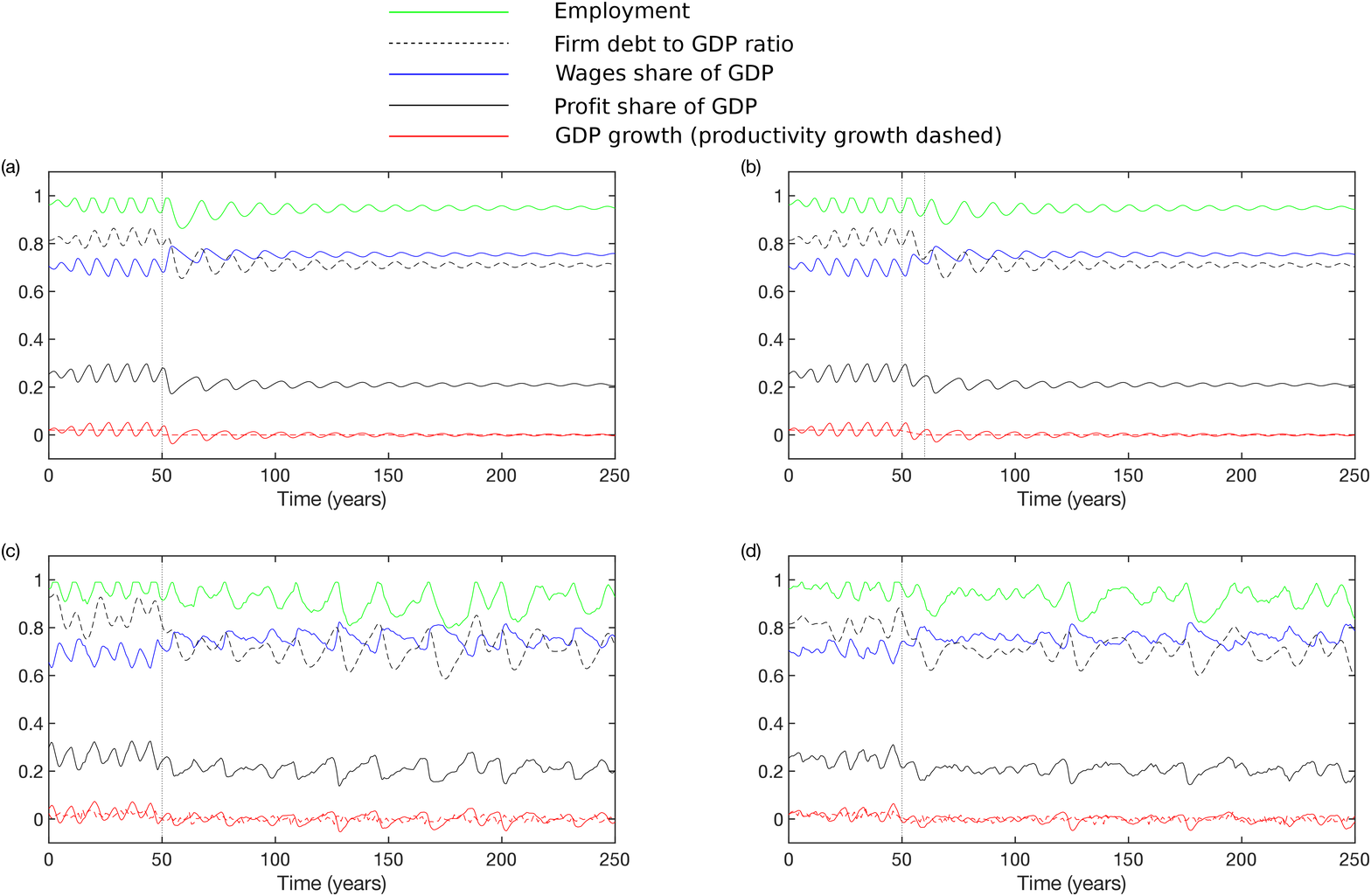}
\end{center}
\caption{Transition from positive growth to zero growth. (a) Constant two percent productivity growth for $t<50$ years, and zero productivity growth thereafter. (b) Constant two percent productivity growth for $t<50$ years; productivity growth decreasing linearly from two percent to zero between $t=50$ years and $t=60$ years; zero productivity growth thereafter. (c, d) Stochastic productivity growth with mean rate of two percent for $t<50$ years and mean rate of zero for $t\geq 50$ years; standard deviation at all times one percent (0.01). These panels show respectively runs in which there isn't and there is a substantial drop in employment shortly after mean growth goes to zero. In all panels the debt behaviour parameters are $\theta_1=\theta_2=0.5$, $\eta_1=3$, $\eta_2=1$, $d_0=0.5$. In all panels the dotted lines show the transition points in productivity growth behaviour. 
} \label{fig:transitiontozerogrowth}
\end{figure*}

Realistically, productivity growth fluctuates, and taking account of this, in Fig.~\ref{fig:stochcases} scenarios with randomly fluctuating productivity growth are shown. In the scenarios in this figure, the productivity growth parameter 
$\alpha$ changes at the beginning of each year. It is independently regenerated each year, from a normal distribution with constant mean (0.02 in the left panels and 0 in the right panels) and a standard deviation of 0.01.\footnote{Such a distribution reflects real data from the UK from the period 1987-2006, during which mean annual productivity growth was $2.13\%$, with a standard deviation of $1.22\%$ (according to the OECD's table at https://data.oecd.org/lprdty/labour-productivity-and-utilisation.htm\#indicator-chart). There was no significant trend in these data (a regression analysis gave $F=0.059$, $p=0.81$) and no significant correlation from one year to the next ($r=0.14$, $p=0.56$).} In the top row of Fig.~\ref{fig:stochcases}, the scenarios from the top row of Fig.~\ref{fig:profitvolcases} are reproduced with such fluctuating productivity growth. Both scenarios remain stable, although there are some sizeable drops in employment for the zero growth case, including one drop down to almost 0.6 during the 250 simulated years. In the middle row of this figure, the scenarios from the bottom row of Fig.~\ref{fig:profitvolcases} are reproduced. Once again, both scenarios remain stable. In this case the fluctuations in employment are comparable for both two percent and zero growth. In the bottom row of Fig.~\ref{fig:stochcases}, it is demonstrated that stochastic productivity growth leads to substantial fluctuations in employment and the profit and wages shares even if debt is held almost constant by the debt behaviour parameters; the scenario $\theta_1=\theta_2=0.1$, $\eta_1=0$, $\eta_2=0$ is plotted. (These scenarios lead to only very small fluctuations in these variables if productivity growth is set constant rather than fluctuating stochastically.) In this case, the fluctuations in employment and profit and wages shares are bigger for the zero growth case. In Fig.~\ref{fig:figappmontecarlo} in Appendix \ref{app:figs}, Monte Carlo simulations are shown for each of the scenarios in Fig.~\ref{fig:stochcases}, namely mean and standard deviation over 1000 implementations are plotted. The Monte Carlo simulations demonstrate that the behaviour seen in the single implementations in Fig.~\ref{fig:stochcases} are typical.

Fig.~\ref{fig:transitiontozerogrowth} shows scenarios for the transition from a positive growth economy to a zero-growth economy, under the debt behaviour parameters $\theta_1=\theta_2=0.5$, $\eta_1=3$, $\eta_2=1$, $d_0=0.5$. Fig.~\ref{fig:transitiontozerogrowth}(a) shows constant two percent productivity growth prior to 50 years, followed by zero productivity growth thereafter. The system remains stable following the end of growth, although there is a temporary substantial drop in employment near the beginning of the zero-growth era, with a low of 0.863. In Fig.~\ref{fig:transitiontozerogrowth}(b), the change in productivity growth is instead implemented gradually, linearly decreasing from 0.02 to 0 over the course of a decade from 50 to 60 years. In this scenario the low in employment is instead 0.881, thus there is not a huge apparent advantage of a gradual over a sudden curtailing of growth. In Fig.~\ref{fig:transitiontozerogrowth}(c,d), two runs are shown in which productivity growth is stochastic as in Fig.~\ref{fig:stochcases}, with mean 0.02 and standard deviation 0.01 before 50 years, and mean 0 and standard deviation 0.01 after 50 years. In Fig.~\ref{fig:transitiontozerogrowth}(c) there is no substantial drop in employment in the period immediately after mean growth goes to zero, while in Fig.~\ref{fig:transitiontozerogrowth}(d) a substantial drop in employment is observed in this period. In the long run, however, in both of these fluctuating productivity growth runs, there are occasional substantial drops in employment after growth has ended. More positively for workers, all of the scenarios in Fig.~\ref{fig:transitiontozerogrowth}, and indeed in the other figures above, show a higher mean wages share of output during zero-growth simulations compared with two percent productivity growth simulations. In summary, the model implies a stable transition to a post-growth economy, albeit with some fluctuations in the level of employment in the absence of an active government.

\subsection{Summary of results}
In summary, we have found that the model can produce stable and unstable runs, both for a positive growth scenario and a zero growth scenario. Further, the simulations suggest that there is no loss of stability when the economy transitions from positive to zero growth. On the contrary, parameters were found that produced a stable run only for the zero growth case and not for the two percent growth case. In general the system is less stable the greater the dependence of the target debt on profit share and instantaneous growth, and the faster the rates of convergence of debt to target debt and target debt to benchmark debt. This is consistent with debt-deflation theories of economic crisis and depression (Fisher, 1932; Fisher, 1933; Minsky, 1986; Minsky, 1992; Keen, 2000). The employment rate was generally less stable for zero growth scenarios than for positive growth scenarios. However, the mean wages share of output was higher for zero growth runs than for positive growth runs with the same parameters.

\section{Discussion}

The question of whether a capitalist economy with interest-bearing debt has a growth imperative has previously received a range of answers from a variety of models pertaining to the stability and viability of states of various variables. The Binswanger (2009) model, which concluded that a desirable zero-growth state was not possible, made some restrictive assumptions, namely that of a constant growth in firm debt at all times, equal to the growth rate of the economy. Further, the wage bill was assumed to be a constant proportion of firm debt.\footnote{For further critique of this model see Richters and Siemoneit (2017a), in particular for the problematic feature that banks are constantly removing money, in the form of retained profits, from the system.} Cahen-Fourot and Lavoie (2016) showed that the Kalecki and Cambridge equations do allow for the possibility of a stationary zero-growth economy, although the analysis just consists of the derivation of a desirable fixed point, and does not address dynamical stability. Berg et al.~(2015) and Jackson and Victor (2015) presented models that have a stationary zero-growth state with some degree of stability against shocks. Further, the analysis in Jackson and Victor (2015) demonstrated a breakdown of stability if the level of investment, or `animal spirits', are too sensitive to current GDP. 

The key novelty in this paper is analysis that compares the relative stability of growth and no-growth scenarios. The model is capable of producing both stable and unstable scenarios, with or without growth, and stability was assessed based on whether or not a given scenario exhibited explosive behaviour; equilibrium (in the form of an attractive fixed point) was not considered necessary for stability. The paper has also provided the first analysis of zero-growth economics on an explicitly Minskyan model; given a level of productivity growth, it is debt behaviour that determines the stability of the model economy. Although the model is very simple, it offers endogenous dynamic wage and employment rates, which enables comparison of the desirability of different scenarios. By contrast, in Berg et al.~(2015), the total wages per output was held constant, and in Jackson and Victor (2015) wage rates were not taken to depend on the level of employment. Another difference between the present model and analysis, and the others above, is the tweaking of a (exogenous) productivity growth input parameter. Jackson and Victor (2016) considered dynamics of the wages share, but there output growth was an input parameter, hence stability of output growth could not be concurrently assessed. 

In several studies, it has been emphasised that consumption must remain high to maintain stability in a zero-growth scenario (Richters and Siemoneit, 2017a). The model here is investment-led, and total consumption of the output produced from full capital utilisation is assumed. Thus, instability through under-consumption cannot occur. The analyses in Richters and Siemoneit (2017a) assume constant rates of consumption out of wealth and income, and show that various models' stable fixed points can be unstable (repulsive) if these parameters are too small. However, in a scenario of recession due to under-consumption, if the rates of consumption again increased, stability of those models would be restored. The model here neglects short-run consumption dynamics, and assumes that in the long-run it is the investment/debt dynamics that determine whether the system exhibits stability or explosive behaviour (crisis).

Another novelty of this paper is the consideration of scenarios in which the productivity growth rate fluctuates around zero. This is more realistic than having it remain constant, and has a profound effect on volatility. Comparing the panels in Figs.~\ref{fig:profitvolcases} and \ref{fig:stochcases} with $\theta=\theta_2=0.75$, $\eta_1=3$, $\eta_2=1$, the former scenario is one of constant equilibrium with constant productivity, and the latter is one of substantial fluctuations in all variables, as a result of simply allowing productivity growth to fluctuate realistically. We have held other variables constant, notably the interest rate, and (implicitly) prices. Future research could explore fluctuations of these parameters, or incorporate dynamic prices as in, say Grasselli and Huu (2015).

How is stability achieved in the model for the zero-growth case? A standard concern is that a positive interest rate leads to an exponentially growing stock of debt in the absence of growth (Douthwaite, 2000; Farley et al., 2013). However, in the many stable scenarios plotted, the profit share (defined as output minus wages minus interest) is positive, and thus firms are (at the aggregate macro level) able to keep up with interest payments to prevent an exponential growth of debt. The desired investment (profit plus new debt) fluctuates, but on average just covers deprecation of the capital stock (average growth of the capital stock is zero). On average, there is precisely zero profit left over after the costs of replenishing the capital stock. Following Richters and Siemoneit (2017a), we consider this a viable scenario; however, firm owners' income must be considered as either negligible or simply part of the wage bill.

In the model, debt dynamics are assumed to be determined by a dynamical target debt that depends on growth and the current profit share. Investment is a `residual variable', that is fixed given the assumptions about debt. The results however are likely to generalise to related models on which debt is the residual, and investment is determined explicitly as a function of profit, debt and growth. In Appendix \ref{sec:generality}, we show for a general class of such models [based on the Keen (1995) model] that there exist, irrespective of whether productivity growth is positive or zero: (i) an economically desirable fixed point (positive wages and employment, and finite debt) that may or may not be attractive; (ii) an attractive fixed point with infinite debt. Behaviour close to the economically desirable fixed point will generally be complex and non-linear; however there is no mathematical reason to expect a demarcation in behaviour between scenarios in which the productivity growth parameter is zero and in which it takes a small positive value e.g.~two percent. We do however find there to be no stable zero growth scenario for the original Keen (1995) model, on which investment decisions are based purely on profit (with no direct dependence on debt).

It is notable that the model shows the wages share to increase when growth decreases to zero (independent of Phillips curve parameters). This is because Piketty's (2014) famous analysis posited the opposite, leading to concerns of there being an incompatibility between sustainability (low growth) and equality (high wages share). Piketty's analysis was neoclassical in nature, and considered only an equilibrium scenario, assuming a constant rate of return on existing wealth, which leads to ever-increasing inequality if output (and the workers' wages share of it) doesn't grow at least as fast as this rate of return. Here, profits, wages and production are placed into a dynamical stock-flow consistent model, and a different conclusion emerges. Our finding here also contrasts somewhat with that of Jackson and Victor (2016). In that paper, the question whether slow growth leads to rising inequality was explored with a stock-flow consistent model with a constant elasticity of substitution production function (a production function associated with neoclassical studies). It was found that only for relatively small values for the elasticity of substitution between labour and capital did inequality not rise for low growth. Here we have utilised the production function $Y=K/\nu$, which is more common in the post-Keynesian literature (Fontana and Sawyer, 2016), and was the one used in the original Keen (1995) Minsky model. Our production function does however make the simplifying assumption of full capacity utilisation. More detailed post-Keynesian models would incorporate incomplete capacity utilisation, see e.g.~Fontana and Sawyer (2016). 

There are obviously many significant omissions to the simple model. As mentioned above, consumer demand dynamics are not modelled, and it is assumed (implicitly) that the supply-driven output can be smoothly absorbed by international markets. The government sector is notably absent. Minsky advocated a big government to stabilise unstable economies (Minsky, 1986; Minsky 1992). Indeed, other studies on similar models have shown that countercyclical government spending can enhance stability (Dafermos, 2017; Costa Lima et al., 2014). The model does not incorporate a financial sector, nor households taking up firm equity, or corporate bonds. Future work will explore the extent to which the modern financial system creates a growth imperative, and in what ways it could be tweaked to improve the viability of low- or no-growth economics.  Further, only a single country is considered. With the profit share decreased for the no-growth compared to growth scenario, in an open-border global economy, capital would flow out of the borders of a no-growth country to a growth country, with potential to cause a crisis from lack of investment (Lawn, 2005, 2011). Further work ought to analyse the extent to which restricting the international mobility of capital would be necessary during the transition to a zero-growth economy. For a recent ecological macroeconomics study with much more detailed modelling see Dafermos et al.~(2017).

The model is macroeconomic in nature and does not address the existence of a growth imperative at the single firm level. Gordon and Rosenthal (2003) analysed this, and concluded there was a growth imperative based on the volatility of profits of typical large firms on the stock market. However, Richters and Siemoneit (2017b) pointed out that several key assumptions of this analysis were unrealistic, for example, the constant investment rates and personal drawing rates. Further, a zero-growth macroeconomic era would likely see reduced volatility of profits, as debt/investment behaviour would likely become less volatile. Thus, there is scope for further combined micro- and macroeconomic analysis of zero growth at the single firm level. Of course in a zero-growth economy, there will still be some businesses that grow alongside others that shrink, and the dynamics of economic transformation and creative destruction will still occur (Jackson, 2009; Malmaeus and Alfredsson, 2017). 

We have not presented arguments for or against desiring zero growth in productivity and/or output, or for the feasibility of long-run zero productivity growth. We have rather attempted to model the consequences of this, should this occur. It is notable that Keynes (1936) envisaged an eventual end to growth. Further, some mainstream economists do now consider that, irrespective of policy, the ``new normal" growth rate is 1\% or lower, possibly due to environmental factors starting to substantially counteract productivity advances from technological development (Malmaeus and Alfredsson, 2017). For recent discussion of prospects for growth and ideologies about growth, see e.g.~Malmaeus and Alfredsson (2017) or Rezai and Stagl (2016). The conclusions of this paper remain valid whether one is interested in the properties of zero-growth economics for reasons of ecological concern (Meadows et al., 1972; Jackson, 2009) or of practical necessity. 

\section{Concluding remarks}
We have analysed the relative stability of positive and zero growth scenarios on a dynamical macroeconomic model with Minskyan features, namely of increasing instability for greater debt behaviour volatility. We found that, all else being equal, zero productivity growth is, if anything, more likely to lead to long-term stability than positive productivity growth, albeit with perhaps a somewhat greater short-term volatility in the oscillatory cycle. Further, according to the model, the end of growth would increase the wages share of output, and hence would not in itself exacerbate inequality. The model contained a basic monetary circuit, and demonstrated the possibility of zero-growth economics with a positive interest rate for loans. Further work will analyse the extent to which other aspects of finance in the modern economy create a growth imperative. 


\section*{Acknowledgements}
ABB is funded by EPSRC grant EP/L005131/1. The Sackler Centre for Consciousness Science is supported by the Dr.~Mortimer and Theresa Sackler Foundation. I am grateful to Yannis Dafermos, Tim Jackson,
Salvador Pueyo and Oliver Richters for feedback on a first draft of this paper. Two anonymous reviewers provided detailed and extremely valuable comments on submitted draft versions of the paper.

\appendix

\section*{Appendix}

\section{Further Simulations} \label{app:figs}

\begin{figure*}
\vspace{1cm}
\begin{center}
\includegraphics[width=0.99\textwidth]{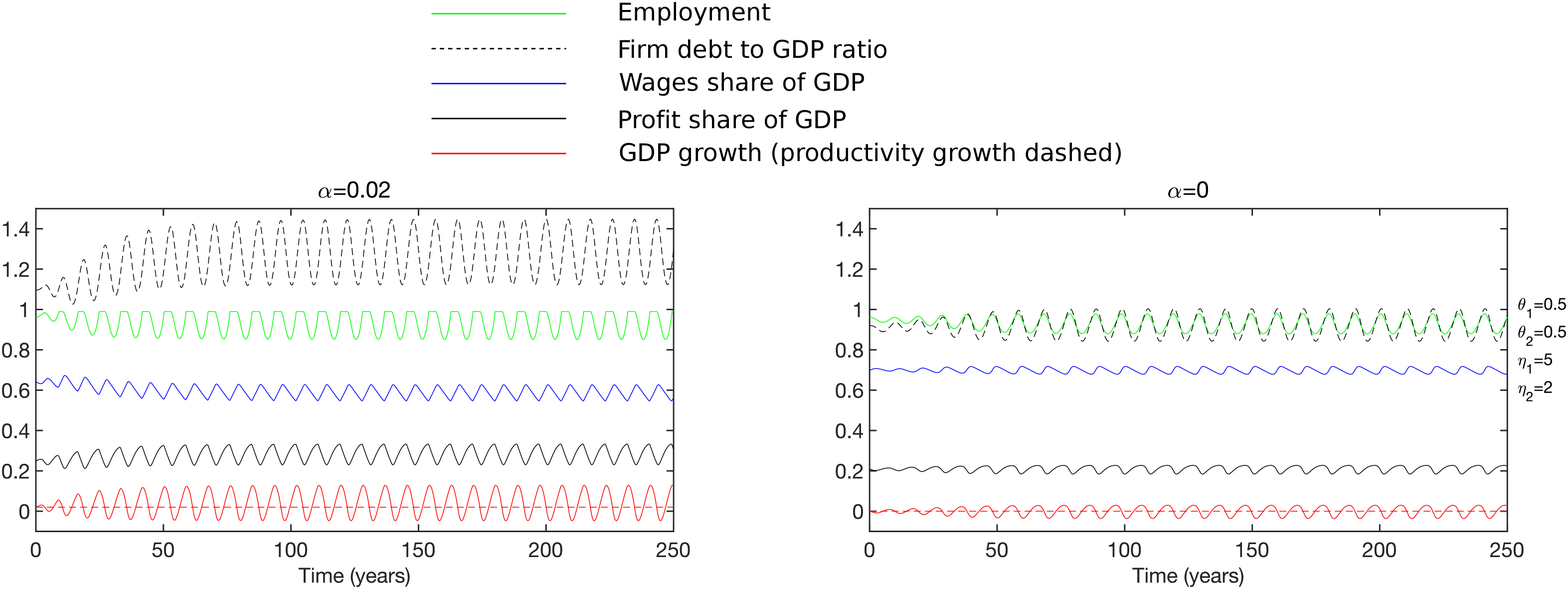}
\end{center}
\caption{Example scenarios with higher interest rate. Simulations identical to those in the second row of Fig.~\ref{fig:profitvolcases}, but with the higher interest rate of $r=0.1$. In both panels $d_0=0.5$. All variables are started at the fixed point, except for $\lambda$ which is initialised at $\bar{\lambda}-0.01$.} \label{fig:figappinterest}
\end{figure*}

\begin{figure*}
\vspace{-2cm}
\begin{center}
\includegraphics[width=0.99\textwidth]{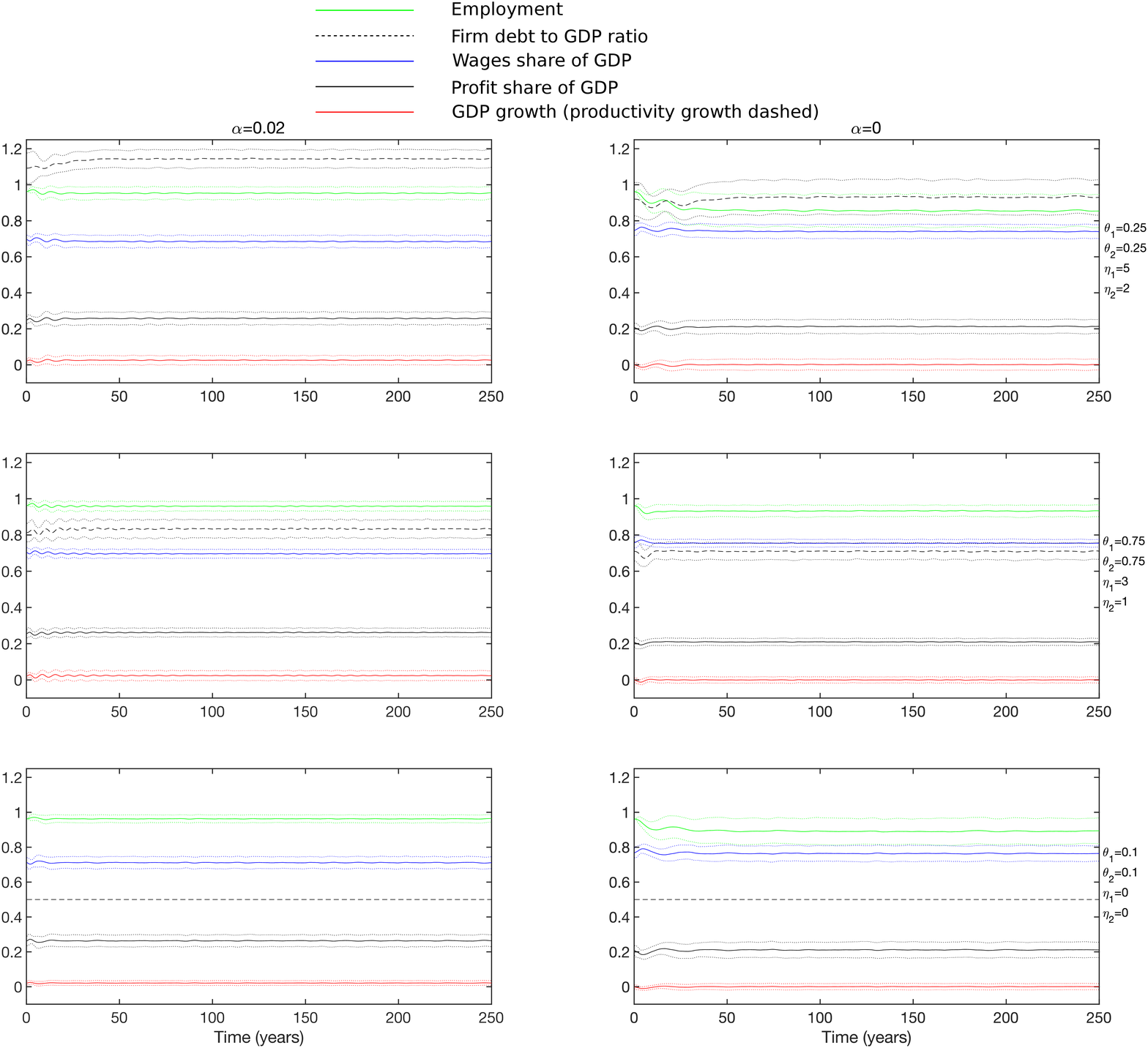}
\end{center}
\caption{Monte Carlo simulations of stochastic productivity growth scenarios. Mean over 1000 implementations of the simulations shown in Fig.~\ref{fig:stochcases}. (Left) Two percent mean productivity growth. (Right) Zero mean productivity growth. In all panels $d_0=0.5$. All variables are started at the fixed point, except for $\lambda$ which is initialised at $\bar{\lambda}-0.01$. Dotted lines show mean plus/minus one standard deviation across implementations. See main text for further details.} \label{fig:figappmontecarlo}
\end{figure*}

Fig. 4 shows example scenarios with a higher interest rate of r = 0.1. Fig. 5 shows mean and
standard deviation over 1000 simulations of each of the stochastic productivity growth scenarios
shown in Fig. 2.

\section{Analysis of fixed point} \label{sec:fixedpoint}
This section presents analysis of the fixed point \eqref{eq:equil1}--\eqref{eq:equil5}. Defining the vector $\boldsymbol{x}=(d_T, d, \lambda, \omega)^{\mathrm{T}}$, the Jacobian $J$ is given by $J_{ij}=:\partial \dot{x}_i/ \partial x_j$, and can be computed at the fixed point as
\be
\bar{J}=\left( \begin{array}{cccc}
\frac{\eta_1 \theta_1\theta_2}{\nu -\bar{d}} - \theta_2 & \frac{\eta_1\theta_2(\alpha-r-\theta_1)}{\nu-\bar{d}}-\eta_2\theta_2 r & 0 & -\frac{\eta_1\theta_2}{\nu-\bar{d}}-\eta_2\theta_2 \\
\theta_1 & -\theta_1 & 0 & 0 \\
\frac{\theta_1\bar{\lambda}}{\nu-\bar{d}} & \frac{\bar{\lambda} (\alpha-r-\theta_1)}{\nu-\bar{d}} & 0 & -\frac{\bar{\lambda}}{\nu-\bar{d}}\\
0 & 0 & \bar{\omega}\Phi^\prime(\bar{\lambda}) & 0 \end{array} \right)\,.
\ee
This can be re-expressed as
\be
\bar{J}=\left( \begin{array}{cccc}
\theta_1  - \theta_2 & -K_2K_5-K_3r& 0 & -K_5-K_3 \\
\theta_1 & -\theta_1 & 0 & 0 \\
\theta_1 K_1 & -K_2K_1 & 0 &-K_1\\
0 & 0 &K_4 & 0 \end{array} \right)\,,
\ee
where
\be
K_1=\bar{\lambda}(\nu-\bar{d})^{-1}\,, \hspace{0.5cm} K_2=r+\theta_1-\alpha\,, \hspace{0.5cm} K_3=\eta_2\theta_2\,, \hspace{0.5cm} K_4=\bar{\omega}\Phi^\prime(\bar{\lambda})\,, \hspace{0.5cm} K_5=\eta_1\theta_2(\nu-\bar{d})^{-1}\,.
\ee
The characteristic polynomial is then
\be
\chi(x)=x^4+p_3x^3+p_2x^2+p_1x+p_0\,,
\ee
where
\begin{eqnarray}
p_3&=&\theta_2+\theta_1(1-K_5)\,,\\
p_2&=&K_1K_4+\theta_1\theta_2-\theta_1^2K_5 + \theta_1(K_2K_5+K_3r)\,\\
p_1&=&K_1K_4[\theta_1(1+K_3)+\theta_2]\,\\
p_0&=&K_1K_4\theta_1[K_3(r+\theta_1-K_2)+\theta_2]\,.
\end{eqnarray}
The Routh-Hurwitz criterion for the fixed point to be attractive is
\be
p_i>0\,, \hspace{0.2cm} \mathrm{for} \hspace{0.2cm}  0 \leq i \leq 3\,, \hspace{0.2cm} p_3p_2>p_1\,, \hspace{0.2cm} \mathrm{and} \hspace{0.2cm} p_3p_2p_1 >p_1^2+p_3^2p_0\,.
\ee
These do not translate into any simple condition on the parameters for the fixed point to be attractive. For the simulations carried out, there were sometimes eigenvalues with positive real parts and sometimes not, indicating that, in the range of parameter space explored, there are cases for which the fixed point is attractive and cases for which it is repulsive. In all simulations there was at least one pair of complex eigenvalues, which explains the observed oscillatory behaviour.

\section{Analysis of similar models with an explicit investment function, and debt as a residual variable} \label{sec:generality}

In this Appendix, we show firstly that the original Keen (1995) model becomes unstable when productivity growth $\alpha \to 0$. Then we show that this does not happen if the model is modified so that the investment function depends explicitly on debt. Further, for this latter scenario, we demonstrate that the fixed point structure is not dependent on whether productivity growth is zero or positive. These models are similar to the main model presented in this paper, however they have debt as the residual variable in the capital account of firms, as opposed to investment.

In the original Keen (1995) model, investment is purely a (increasing) function of the profit share,
\be
I=\frac{\kappa(\pi)}{\nu} K\,.
\ee
This leads to the system of equations (Grasselli and Costa Lima, 2012):
\begin{eqnarray}
\dot{\omega}&=&\omega[\Phi(\lambda)-\alpha]\,,\\
\dot{\lambda}&=&\lambda(g-\alpha)\,,\\
\dot{d}&=&\kappa(\pi)-\pi-gd\,,\\
g&=&\frac{\kappa(\pi)}{\nu}-\delta\,.
\end{eqnarray}
Assuming $\kappa$ is such that there exists a $\bar{\pi}\in(0,1)$ for which
\be
\kappa(\bar{\pi})=\nu(\alpha+\delta)\,,
\ee
there is a single economically desirable fixed point ($\lambda>0$, $\omega>0$, $d$ finite), given by
\begin{eqnarray}
\bar{\lambda}&=&\Phi^{-1}(\alpha)\,,\\
\bar{d}&=&\frac{\kappa(\bar{\pi})-\bar{\pi}}{\alpha}\,, \label{eq:keendiverge}\\
\bar{\omega}&=&1-\bar{\pi}-r\bar{d}\,.
\end{eqnarray}
For stable scenarios, the system oscillates close to this fixed point (Grasselli and Costa Lima, 2012). On this model, stability can not be maintained as productivity growth $\alpha \to 0$, because \eqref{eq:keendiverge} implies that the debt at the fixed point goes to infinity. 

We now consider modification to this model so that the investment function has an additional direct dependence on debt, i.e.~we replace $\kappa(\pi)$ above with $\kappa(\pi,d)$.\footnote{We don't consider $\kappa$ here to depend on recent (productivity or output) growth. It is straightforward to see that the results here generalise to such cases; the crucial factor is that $\kappa$ has a direct dependence on debt, and that there is a solution to the equivalent of  \eqref{eq:kappapid}, with all variables taking economically desirable values.}   Then \eqref{eq:keendiverge} becomes
\be
\kappa(\bar{\pi},\bar{d})=\bar{\pi}+\alpha \bar{d}\,. \label{eq:kappapid}
\ee
There is a broad space of functions $\kappa(\pi,d)$ for which a solution to this equation exists for both $\alpha=0$ and small positive values of $\alpha$. Thus, in general there will be an economically desirable fixed point for both $\alpha=0$ and, say $\alpha=0.02$.  Such a system also has a fixed point with infinite debt, $(\omega,\lambda,d)=(0,0,\infty)$, and a possible cause of run-away behaviour is this fixed point being attractive. The existence of this fixed point is easily verified by considering the transformed system with $d$ replaced by $u=:1/d$ [again following Grasselli and Lima (2012)],
\begin{eqnarray}
\dot{\omega}&=&\omega[\Phi(\lambda)-\alpha]\,,\\
\dot{\lambda}&=&\lambda(g-\alpha)\,,\\
\dot{u}&=&[u(1-\omega)-r+g-u\kappa(\pi,u^{-1})]u\,.
\end{eqnarray}
The Jacobian at this fixed point $(\omega,\lambda,u)=(0,0,0)$ is given by
\be
J(0,0,0)=\left( \begin{array}{ccc}
\Phi(0)-\alpha & 0& 0  \\
0 & \frac{\kappa(-\infty,\infty)-\nu(\alpha+\delta)}{\nu} & 0  \\
0 & 0 &\frac{\kappa(-\infty,\infty)-\nu(r+\delta)}{\nu} \end{array} \right)\,.
\ee
This is diagonal, and hence the condition for the (infinite debt) fixed point to be attractive is that the diagonal components are all negative. Given the assumption $\Phi(0)<0$, and given that $\alpha$ and $r$ are greater or equal to zero, these components are indeed negative if 
\be
\frac{\kappa(-\infty,\infty)}{\nu}<\delta\,,
\ee
that is if the rate of investment is less than the rate of depreciation of capital in the worst case limit scenario of infinite loss (negative profit share) and infinite debt. This would normally be assumed to be the case. Thus, the fixed point at infinite debt is typically stable irrespective of the value of productivity growth $\alpha$. We conclude that the fixed point structure is not dependent on whether productivity growth is zero or positive. Hence, from this analysis, there is no reason that a generalised Keen model, with an investment function that depends explicitly on debt, should be any more or less stable when productivity growth is zero, as opposed to say two percent. 

\section{Supplementary data}
Matlab code included as Supplementary Material can be found online at:

http://www.sciencedirect.com/science/article/pii/S0921800917306869 

 \section*{References}

Berg, M., Hartley, M., Richters, O., 2015. A stock-flow consistent input-output model with applications to energy price shocks, interest rates, and heat emissions. New J.~Phys.~17(1), 015011. \\

\noindent Binswanger, M., 2009. Is there a growth imperative in capitalist economies? A circular flow perspective. J.~Post Keynes.~Econ.~31, 707-727.\\

\noindent Cahen-Fourot, L., Lavoie, M., 2016. Ecological monetary economics: A post-Keynesian critique. Ecol.~Econ.~126, 163-168.\\

\noindent Costa Lima, B., Grasselli, M.R., Wang, X.-S., Wu. J., 2014. Destabilizing a stable crisis: Employment persistence and government intervention in macroeconomics. Structural Change and Economic Dynamics 30, 30-51.\\

\noindent Dafermos, Y., 2017. Debt cycles, instability and fiscal rules: A Godley-Minsky model. Camb.~J.~Econ.~in press. Available from: http://eprints.uwe.ac.uk/26694.\\

\noindent Dafermos, Y., Nikolaidi, M., Galanis, G., 2017. A stock-flow-fund ecological macroeconomic model. Ecol.~Econ.~131, 191-207.\\

\noindent Douthwaite, R., 2000. The Ecology of Money. Green Books, Devon.\\

\noindent Farley, J., Burke, M., Flomenhoft, G., Kelly, B., Murray, D., Posner, S., Putnam, M., Scanlan, A., Witham, A., 2013. Monetary and fiscal policies for a finite planet. Sustainability 5, 2802-2826.\\

\noindent Fisher, I., 1932. Booms and depressions. Adelphi, New York.\\

\noindent Fisher, I., 1933. The debt-deflation theory of great depressions. Econometrica 1, 337-357.\\ 

\noindent Fontana, G., Sawyer, M., 2016. Towards post-Keynesian ecological macroeconomics. Ecol.~Econ.~21, 186-195.\\

\noindent Goodwin, R.M., 1967. A growth cycle. In: Feinstein, C.H. (ed.) Socialism, Capitalism and Economic Growth, pp. 54-58. Cambridge University Press, Cambridge.\\

\noindent Gordon, M.J., Rosenthal, J.S., 2003. Capitalism's growth imperative. Camb.~J.~Econ.~27, 25-48.\\

\noindent Grasselli, M.R., Costa Lima, B., 2012. An analysis of the Keen model for credit expansion, asset price bubbles and financial fragility, Mathematics and Financial Economics, 6 (3), 191-210.\\

\noindent Grasselli, M.R., Huu, A.N., 2015. Inflation and Speculation in a Dynamic Macroeconomic Model. J.~Risk Financial Manag.~8, 285-310.\\

\noindent Hardt, L., O'Neill, D.W., 2017. Ecological macroeconomics models: Assessing current developments. Ecol.~Econ.~134, 198-211.\\

\noindent Jackson, T., 2009. Prosperity Without Growth - Economics for a Finite Planet. Routledge, London.\\

\noindent Jackson, T., Victor, P.A., 2015. Does credit create a `growth imperative'? A quasi-stationary economy with interest-bearing debt. Ecol.~Econ.~120, 32-48.\\

\noindent Jackson, T., Victor, P.A., 2016. Does slow growth lead to rising inequality? Some theoretical reflections and numerical simulations. Ecol.~Econ.~121, 206-219.\\

\noindent Keen, S., 1995. Finance and economic breakdown: modeling Minsky's ``Financial Instability Hypothesis'' J.~Post Keynes.~Econ.~17(4), 607-635.\\

\noindent Keen, S., 2000. The nonlinear economics of debt deflation. In: Barnett, W.A., Chiarella, C., Keen, S., Marks, R., Schnabl, H. (eds.) Commerce, Complexity, and Evolution: Topics in Economics, Finance, Marketing, and Management, pp. 83-110. Cambridge University Press, Cambridge.\\

\noindent Keen, S., 2011. Debunking Economics - Revised and Expanded Edition. Zed Books, London and New York.\\

\noindent Keen, S., 2013. A monetary Minsky model of the Great Moderation and the Great Recession. Journal of Economic Behavior \& Organization, 86, 221-235.\\

\noindent Keynes, J.M., 1936. The General Theory of Employment, Interest, and Money. McMillan, London.\\

\noindent Lawn, P., 2005. Is a democratic-capitalist system compatible with a low-growth or steady-state economy? Soc.~Econ.~Rev.~3, 209-232.\\

\noindent Lawn, P., 2011. Is steady-state capitalism viable? A review of the issues and an answer in the affirmative. Ann.~N.~Y.~Acad.~Sci.~1219, 1-25.\\

\noindent Malmaeus, J.M., Alfredsson, E.C., 2017. Potential consequences on the economy of low or no growth - short and long term perspectives. Ecol.~Econ.~134, 57-64.\\

\noindent Meadows, D.H., Meadows, D.L., Randers, J., Behrens, W.W., 1972. The Limits to Growth. Universe Books, New York.\\

\noindent Minsky, H.P., 1986. Stabilizing an Unstable Economy. Mc Graw Hill, New York.\\

\noindent Minsky, H.P., 1992. The financial instability hypothesis. Working Paper 74. The Jerome Levy Economics Institute of Bard College. Annandale-on-Hudson, NY. Available from:\\ http://www.levyinstitute.org/pubs/wp74.pdf.\\

\noindent Nikolaidi, M., Stockhammer, E., 2017. Minsky Models. A structured survey. Working paper PKWP1706. Post-Keynesian Economics Study Group.\\

\noindent Piketty, T., 2014. Capital in the 21st century. Harvard University Press, Harvard.\\

\noindent Rezai, A., Stagl, S., 2016. Ecological macroeconomics: introduction and review. Ecol.~Econ.~121, 181-185.\\

\noindent Richters, O., Siemoneit, A., 2017a. Consistency and stability analysis of models of a monetary growth imperative. Ecol.~Econ.~136, 114-125.\\

\noindent Richters, O., Siemnoneit, A., 2017b. Fear of stagnation? A review on growth imperatives. Vereinigung f\"ur \"Okologische \"Okonomie Discussion Paper, 6/2017.\\

\noindent Rosenbaum, E., 2015. Zero growth and structural change in a post Keynesian growth model. J.~Post~Keynes.~Econ.~37, 623-647.\\

\end{document}